\documentclass{elsart}

\usepackage{graphics,epsfig}
\usepackage{amssymb}

\def\lsim{\lower.5ex\hbox{$\; \buildrel < \over \sim \;$}}
\def\gsim{\lower.5ex\hbox{$\; \buildrel > \over \sim \;$}}

\def\t{\ifmmode {\tau} \else $\tau$ \fi}

\def\cm{\ifmmode {\rm cm}^{-1} \else cm$^{-1}$ \fi}
\def\s{\ifmmode {\rm s}^{-1} \else s$^{-1}$ \fi}
\def\cc{\ifmmode {\rm cm}^{-3} \else cm$^{-3}$ \fi}
\def\cs{\ifmmode {\rm cm}^{-2} \else cm$^{-2}$ \fi}
\def\g{\ifmmode \gamma \else $\gamma$\fi}
\def\gs{\ifmmode \gamma \else $\gamma~$\fi}
\def\l{\ifmmode \lambda \else $\lambda$\fi}

\def\t{\ifmmode \tau \else $\tau$\fi}
\def\G{\ifmmode \Gamma \else $\Gamma$\fi}
\def\Gt{\ifmmode \tilde{\Gamma} \else $\tilde{\Gamma}$\fi}

\def\kms{\ifmmode {\rm km\ s}^{-1} \else km s$^{-1}$\fi}

\def\ks{``knee"~}
\def\k{``knee"}

\begin{document}

\begin{frontmatter}

\title{Cosmic Rays and Large Extra Dimensions 
}
\author{D. Kazanas}
\address{
LHEA, NASA/GSFC Code 661, Greenbelt, MD 20771
}
\author{A. Nicolaidis}
\address{
Department of Theoretical Physics, University of Thessaloniki,
Thessaloniki 54006, Greece
}
\begin{abstract}
\noindent 
We have proposed that the cosmic ray spectrum \k, the steepening
of the cosmic ray spectrum at energy $E \gsim 10^{15.5}$ eV, is due 
to ``new physics", namely new interactions at TeV cm energies which 
produce particles undetected by the experimental apparatus. In this
letter we examine specifically the possibility that this interaction is 
low scale gravity. We consider that the graviton propagates, besides 
the usual four dimensions, into an additional $\delta$, compactified, large 
dimensions and we estimate the graviton production in $p\,p$ collisions
in the high energy approximation where graviton emission is factorized.
We find that the cross section for graviton production rises as fast as 
$(\sqrt{s}/M_f)^{2+\delta}$, where $M_f$ is the fundamental scale of 
gravity in $4+\delta$ dimensions, and that the distribution of radiating a 
fraction $y$ of the initial particle's energy into gravitational energy
(which goes undetected) behaves as $\delta y^{\delta -1}$. The missing
energy leads to an underestimate of the true energy and generates a
break in the {\sl inferred} cosmic ray spectrum (the \k). By fitting 
the cosmic ray spectrum data we deduce that the favorite values for the
parameters of the theory are $M_f \sim 8$ TeV and $\delta =4$.

\end{abstract}

\begin{keyword}

\PACS {95.30Cq},~{96.40De},~{98.7Sa},~{12.60.-i}

\end{keyword}
\end{frontmatter}



\label{sec:intro}

Cosmic rays (CR) is a subject almost a century old; despite 
the amount of knowledge accumulated since their discovery, 
there is a number of outstanding issues concerning their origin, 
acceleration and composition. One of the most interesting features is the 
breadth of their energy spectrum which extends over 11 orders of magnitude
up to and beyond $10^{20}$ eV (see \cite{R1} for a recent review). 

The CR spectrum can be described as a broken power law $E^{-\g}$, with 
\gs obtaining three different values in three different energy regimes:
In the range $10^9-10^{15.5}$ eV, $\g \simeq 2.75$. At energies greater
than $10^{15.5}$, the \k, the spectrum steepens to $\g \simeq 3$ out 
to an energy $E \sim 10^{18}$ eV (the ``ankle"), at which point it
becomes flatter again, $\g \simeq 2.2 - 2.5$ and extends to $\sim 
10^{20.5}$ eV, beyond which the flux is too small to give a meaningful
flux even with the largest current detectors. 

The currently most popular models for cosmic ray acceleration (supernova 
shocks) can account for the CR spectra up to energies $~10^{14}Z$ eV
only ($Z$ is the corresponding nuclear charge). Measuring the composition
and understanding the origin of the cosmic ray spectrum above these 
energies remains a challenge both experimentally and theoretically.
Of particular interest in this respect is the origin of the 
steepening of the CR spectrum at the \k: As we suggested in 
earlier work \cite{R2}, this (in fact any) spectral steepening cannot be 
accounted for by considering the spectrum to be the sum of two 
independent CR components, presumably the output of two different
kinds of CR sources (an explanation that would be valid for a 
flattening of the spectrum such as that observed at the ``ankle"). 
Furthermore, we suggested in the above reference that the simplest
explanation of the CR spectrum break at the \ks is that it is due
to a novel channel in the physics of high energy $pp$-collisions, 
as proposed also by others in the past \cite{nikol93}. 
We also noted there that the cm energy corresponding to the 
\ks is a few TeV, a scale tantalizingly close to 
the energy at which the emergence of ``new physics" is anticipated 
on very general grounds.  We advanced the idea that the new 
interaction, operative at and above these energies, results
in particles that are not detected by the associated experimental 
devices, leading to an underestimate of the incident particle's 
energy for energies above that of the \k. As a result, a cosmic ray
spectrum which is a {\sl single} power law in energy, will develop
an increase in its slope (a \k) at the energy at which the new 
interaction turns-on, with the spectrum reverting to its original
slope when saturation occurs.

The main motivation for introducing new physics comes from the need 
to provide a unified theory in which two disparate scales, i.e. the 
electroweak $M_W \sim 100$ GeV and the Planck scale 
$M_P \sim 10^{19}$ GeV can coexist (hierarchy problem). 
Recently a novel approach has been proposed for resolving the 
hierarchy problem \cite{R3}. Specifically, it has been suggested
that our four dimensional world is embedded in a higher dimensional
space with $D$ dimensions of which $\delta$ dimensions are compactified
with a relatively large radius. While the Standard Model (SM) fields 
live on the 4-dimensional world (brane), the gravitons can propagate 
freely in the higher dimensional space (bulk). The fundamental scale 
$M_f$ of gravity in $D$ dimensions is related to the observed 
4-dimensional Newton constant $G_N$ by
\begin{equation}
G_N = \frac{1}{V_{\delta}} \left(\frac{1}{M_f} \right)^{(2+\delta)}
\label{gnewt}
\end{equation}
where $V_{\delta}$ is the volume of the extra space. For a torus
configuration
\begin{equation}
V_{\delta} = (2 \pi R)^{\delta}
\end{equation}
with $R$ the common radius of the large extra dimensions. For a given $\delta$, 
a sufficiently large radius $R$ can then reduce the fundamental scale of
gravity $M_f$ to TeV energies which are not too different from $M_W$,
 thereby resolving the hierarchy problem. 

The $D-$dimensional graviton, when reduced to $4-$dimensions, gives
rise to spin 2 particles (the massless graviton and its massive 
Kaluza-Klein excitations), spin 1 particles and spin zero particles. 
The vector particles do not couple to the energy momentum tensor and
are ignored, while the scalar particles couple to the trace of the 
energy momentum tensor, giving negligible contributions to the high 
energy regime we are interested in \cite{R4,R5}. Thus we take into 
account only the production of spin 2 gravitons in the $pp$-collisions 
we consider.

We study the $pp \rightarrow pp$ collision at the cm system, where 
each particle carries energy $\sqrt{s}/2$. At high energies the 
dominant contribution to graviton ($h$) production ($pp \rightarrow
pph$), originates from collinear bremsstrahlung of gravitons from 
each external line. In this configuration, graviton emission is 
factorized  as the probability that a proton (incoming or outgoing)
loses a fraction $y$ of its energy via graviton radiation. The 
Kaluza-Klein (KK) excitations of the graviton have the same couplings
to ordinary fields as their massless zero-mode, i.e. they couple 
to the $4-$momentum of the fermion they are attached. Within this 
reasonable approximation of factorized emission, we obtain that the
cross-section for graviton production in $pp$ collisions, with the 
graviton carrying energy $\epsilon$, is given by
\begin{equation}
\frac{d \, \sigma_h}{d \, \epsilon}(pp \rightarrow pph)
= \sigma_0 \, \frac{4 G_N}{\pi} \, s \; log \left(\frac{s}{m_p^2}
\right) \, \frac{g(\epsilon)}{\epsilon}
\label{xsec}
\end{equation}
where $\sigma_0$ is the $pp \rightarrow pp$ cross-section
(it rises slowly with energy), $m_p$ is the proton mass
and $g(\epsilon)$ is the multiplicity factor counting the 
number of KK gravitons contributing to the process. The
mass of each KK mode corresponds to the modulus of its 
momentum in the direction transverse to the brane
\begin{equation}
m_n^2 = \frac{ \bf n^2}{R^2}
\end{equation}
with ${\bf n} = (n_1, n_2, ..., n_{\delta}$). At fixed $\epsilon$,
all gravitons with $m_n \le \epsilon$ contribute, their number
being the volume of a $\delta-$dimensional sphere with radius
$n_{\rm max} = R \epsilon$. Thus 
\begin{equation}
g(\epsilon) = \frac{2 \pi^{\delta/2}}{\delta \G(\delta/2)}
(R \epsilon)^{\delta}
\label{multipl}
\end{equation}
Substituting expression (\ref{multipl}) into eq. (\ref{xsec}),
using relation (\ref{gnewt}), absorbing constants and 
$\delta-$dependent factors into a rescaled $M_f$, we find that
the cross-section for a fractional energy loss $y~
(y=2\,\epsilon/\sqrt{s})$ is 
\begin{equation}
\frac{d \, \sigma_h}{d \, y}(pp \rightarrow pph)
= \sigma_0 F(s) f(y)
\label{xsec2}
\end{equation}
with 
\begin{equation}
F(s) = \left(\frac{\sqrt{s}}{M_f}\right)^{2+\delta}
log \left( \frac{s}{m_p^2}\right)
\end{equation}
\begin{equation}
f(y) = \delta y^{\delta -1}
\end{equation}

We observe that the cross-section for graviton emission rises fast, 
as a power of     the energy (i.e. $\propto (\sqrt{s}/{M_f})^{2+\delta}$)
and while it is unimportant at energies below the scale $M_f$, it is
the dominant process at and just above $M_f$. The distribution of
the energy radiated into  massless gravitons is represented by
the well-known infrared  $1/y$ behavior, but the large number of 
KK modes converts this distribution into $y^{\delta - 1}$. We 
emphasize again that within our approximative scheme we pick-up 
the leading contributions originating from collinear, soft, gravitational
radiation. For the purposes of the present investigation we consider
this approximation as well justified.

\begin{figure}[hbt]
\centerline{\epsfig
{file=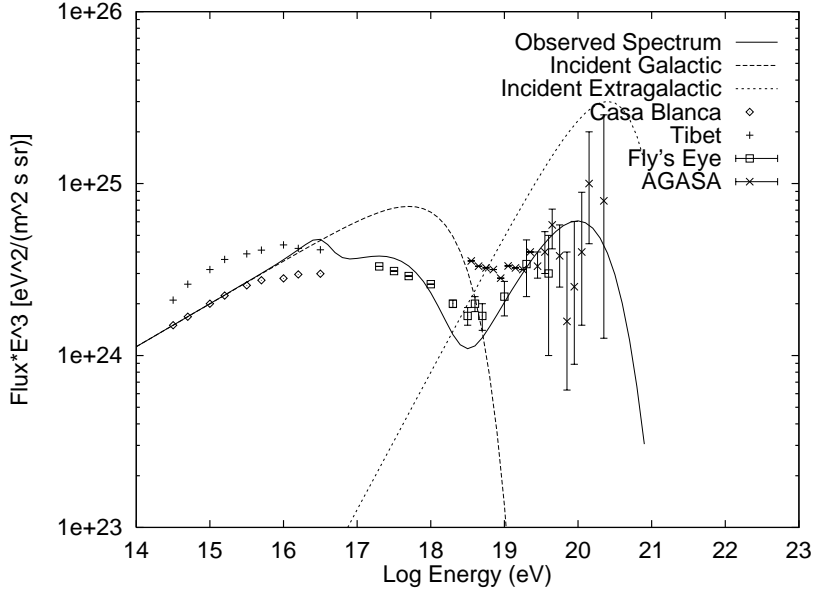,width=.8\textwidth,angle=0}}
\caption{The cosmic ray spectrum $f(E)$ multiplied by $E^3$ for 
$E>10^{14}$ eV. Long and
short dashed lines are respectively the incident galactic and 
extragalactic components. Solid line is the combined spectrum expected 
to be measured for $M_f = 8\; {\rm TeV},~\delta=4$.
$+$'s are the Tibet data, diamonds the Casa Blanca data, squares the 
Fly's Eye data and $\times$'s the AGASA data.}
\label{f2}
\end{figure}

In the lab system a cosmic ray particle of high energy hits a nucleon at 
rest in the Earth's atmosphere. For a fractional energy loss $y$ in the
cm system, in the lab system the corresponding event of total energy
$E^{\prime}$ will be registered at the detector at an energy $E = 
E^{\prime}(1-y/2)^2$. Therefore, for a cosmic ray intensity $I(E)$,
gravitational radiation will induce an {\sl inferred} intensity 
$N(E)$ of the form
\begin{equation}
N(E) = \int d \, E^{\prime} \int_0^1 d\,y f(y) I(E^{\prime}) 
\delta (E - E^{\prime}(1-y/2)^2))
\label{shift}
\end{equation}

The cosmic ray interactions will proceed either through the standard
channels with probability $P_0(E)$, or through the production of 
gravitons with probability $P_h(E)$. We deduce that the observed 
cosmic ray intensity $I_{ob}(E)$ at an energy $E$ will be the sum of 
these two processes, i.e. 
\begin{equation}
I_{ob}(E) = P_0 I(E) + P_h N(E)
\end{equation}
where
\begin{eqnarray}
P_0(E) & =  \frac{\sigma_0}{\sigma_0 + \sigma_h} & =  \frac{1}{1+F}\\
P_h(E) & =  \frac{\sigma_h}{\sigma_0 + \sigma_h} & =  \frac{F}{1+F}
\end{eqnarray}

We assume that the incident cosmic ray spectrum $I(E)$ is composed of two
components, the galactic component ($I_G(E)$) and the extragalactic 
component ($I_E(E)$) each of which is assumed to be a power law of indices
$\g = 2.75 ~{\rm and}~ 2.2$ respectively, with the galactic component 
being dominant at energies $E \lsim 10^{18.5}$ eV and the extragalactic one
at higher energies as discussed in \cite{R2} (i.e. $I_G(E) \propto 
E^{-2.75} \, exp(-E/E_0), ~ E_0 \sim 10^{18.5}$ eV and $I_E(E)\propto E^{-2.2}
exp(-E/E_1), ~ E_1 \sim 10^{20.5}$ eV). 
According to our suggestion, the observed spectrum $I_{ob}(E)$ results from a
migration of high energy points of energy $E^{\prime}$ to lower energy
$E$ [see Eq. (\ref{shift})]. Therefore, the inferred spectrum is a 
sensitive function of the input spectrum $I(E)$. Indeed the resulting 
spectrum, shown in fig. 1, depends both on the parameters of the theory
($M_f, \delta$) and the parameters of the input spectrum, notably the 
value of the galactic cut-off energy $E_0$ . Rather than varying freely the
parameters of the input spectrum, we have chosen the 
values adopted in our previous work \cite{R2} for the reasons explained
there. The uncertainty in the experimental data, exemplified by the 
disagreement between two experiments in each energy range  (i.e. 
Tibet - CASA BLANCA for $E = 10^{14.5}-10^{16.5}$ eV , Fly's Eye - 
AGASA for $E > 10^{17.5}$ eV) presented in figure 1, does not 
allow a high significance determination of the parameters of the 
theory. We found however that the data favor the values $M_f \simeq
8$ TeV and $\delta=4$ for the theoretical  parameters, with the values 
$\delta = 3, ~5$ being of lower significance as it was determined by the
numerical study.  Our deduced values respect all available 
bounds on $M_f$ and $\delta$. The most severe bounds come from the 
observations of SN 1987a, which constrain the rate of energy loss due
to new particles \cite{sn}. Detailed caclulations \cite{snraffelt}
provide the limit $M_f \gsim 1$ TeV for $\delta =4$.

Cosmic accelerators accelerate particles (the cosmic rays) 
to energies much higher than those achieved at terrestrial laboratories.
As such they present us with the opportunity of detecting the
signatures of processes operative at energies not yet accessible 
at terrestrical accelerators. As mentioned in the beginning of the 
present note, the lowering of the fundamental Planck scale $M_f$ 
to TeV energies, implies that the gravitational interaction becomes 
strong at these energies. Any scattering process 
at cm energies $E \sim M_f$ should be accompanied by abundant graviton 
production. Since the energy of the cosmic rays can by far exceed this 
threshold domain, graviton emission should be pre-eminent in CR 
interactions. The essence of our proposal is that the signature of 
emission of these gravitons, which go undetected in the cosmic ray air 
shower arrays, is the observed ``knee structure" in the CR spectrum. 

We would like to conclude with a few general remarks.
How reliable is our calculation? It is obvious that our cross section for
graviton emission rises very fast and at some energy unitarity will be
violated. The KK formulation we employed is an effective theory one
and at some energy $M_s$ ($M_s$ is the string scale with $M_s>M_f$)
we have to resort to the underlying theory, a string theory. With 
gravity dominant at TeV scales, gravitational radiation is only one 
of the possible manifestations. Exchange of virtual KK gravitons 
enhances the cross-sections \cite {R8} and this enhancement might help 
to resolve the paradox with the CR events above the GZK cut-off \cite{greis66}.
It has been suggested also that black holes may in fact be produced
in $pp$ collisions \cite{R9}. Their subsequent decay through the 
Hawking radiation should lead to events of high multiplicity 
and large sphericity at the cm. It is of interest that detailed analysis of 
the EAS data suggest an apparent, very sharp change in the CR composition
to almost exculsively Fe just above the \ks (see fig. 5 of \cite{casab}).
This apparent change is qualitatively of the form expected by a 
sharp increase in the interaction cross section and an ensuing 
dispersion of the available energy to a large number of secondary
particles. The process described just above might then provide an 
account of this fact, though we do not believe that this interpretation
can be considered at this point as unique. 
Finally, on the cosmic ray physics side, our proposal 
implies that the galactic
CR sources produce power law spectra extending to the ``ankle"
rather than the \ks (as thought on the basis of acceleration in SN 
shocks). Given that the latter do constitute CR sources to these energies
our arguments imply the presence of two independent galactic CR 
components: one due to SN shocks, while clues of the origin of the second, being 
the observed anisotropy of CR at $E \sim 10^{18}$ eV toward the 
galactic center \cite{agasa}. It appears that cosmic rays rather than being
the messengers, constitute the message itself.

DK  would like to thank Frank Jones and Bob 
Streitmatter for a number of discussions, comments, criticism 
and encouragement. AN would like to acknowlegde useful discussions 
with Ignatios Antoniadis and Costas Counnas. Part of this 
work was presented by AN at the Paris workshop on 
Physics and Astrophysics of Extra Dimensions (May 29 2001) 
and at the NESTOR Institute, Pylos (June 28 2001).

\end{document}